\documentclass[a4paper,11pt]{article}
\usepackage{pos}
\usepackage{epsf,amsmath,graphicx,scalefnt,rotating,enumitem,fancyhdr}

\usepackage[english]{babel}
\usepackage[final]{showkeys}
\usepackage{slashed}
\usepackage{dsfont}
\usepackage[acronym]{glossaries}
\makeglossaries
\glsdisablehyper
\let\oldglsentryshort\glsentryshort
\renewcommand{\glsentryshort}[1]{\abbrev{\oldglsentryshort{#1}}}
\usepackage{filemod}
\usepackage[capitalize]{cleveref}

\usepackage{comment} 

\newcounter{notecount}

\newcommand{\flowt}{t}
\newcommand{\op}{O}

\newcommand{\lhs}{l.h.s.}
\newcommand{\rhs}{r.h.s.}

\newcommand{\citere}[1]{Ref.\,\cite{#1}}
\newcommand{\citeres}[1]{Refs.\,\cite{#1}}

\newcommand{\abbrev}[1]{{\scalefont{.9}#1}}
\newcommand{\EulerGamma}{\gamma_\text{E}}

\newcommand{\bare}{\text{\abbrev{B}}}
\newcommand{\ep}{\epsilon}

\newcommand{\dd}{\mathrm{d}}
\newcommand{\deriv}[3]{\frac{\partial\ifthenelse{\equal{#1}{}}{}{^{#1}}%
    #2}{\partial #3\ifthenelse{\equal{#1}{}}{}{^{#1}}}}
\newcommand{\dderiv}[3]{\frac{\dd\ifthenelse{\equal{#1}{}}{}{^{#1}}%
    #2}{\dd #3\ifthenelse{\equal{#1}{}}{}{^{#1}}}}
\newcommand{\order}[1]{\mathcal{O}(#1)}

\newcommand{\msbar}{\ensuremath{\overline{\mbox{\abbrev{MS}}}}}

\newacronym[\glslongpluralkey={degrees of freedom},%
  shortplural=d.o.f.]{dof}{d.o.f.}{degree of freedom}

\newacronym[\glslongpluralkey={renormalization-group equations},%
  shortplural=\abbrev{RGE}s]{RGE}{RGE}{renormalization-group equation}

\newacronym{FVS}{FVS}{fixed-volume scheme}
\newcommand{\fvs}{\gls{FVS}}
\newacronym[\glslongpluralkey={renormalization groups},%
  shortplural=\abbrev{RG}s]{RG}{RG}{renormalization group}
\newcommand{\rg}{\gls{RG}}
\newacronym{LSZ}{LSZ}{Lehmann-Symanzik-Zimmermann}

\newacronym{SFTX}{SFTX}{short-flow-time expansion}
\newcommand{\sftx}{\gls{SFTX}}
\newacronym{OPE}{OPE}{operator-product expansion}
\newcommand{\ope}{\gls{OPE}}
\newacronym{GF}{GF}{gradient flow}
\newcommand{\gf}{\gls{GF}}
\newacronym{RF}{RF}{Ricci flow}
\newcommand{\rf}{\gls{RF}}
\newacronym[\glslongpluralkey={equation-of-motions},%
  shortplural=\abbrev{EoM}s]{EoM}{EoM}{equation-of-motion}

\newacronym[\glslongpluralkey={Effective Field Theories},%
  shortplural=\abbrev{EFT}s]{EFT}{EFT}{Effective Field Theory}

\newacronym{QED}{QED}{quantum electrodynamics}
\newcommand{\QED}{\gls{QED}}
\newacronym{SSB}{SSB}{spontaneous symmetry breaking}

\newacronym[\glslongpluralkey={vacuum expectation values},%
  shortplural=\abbrev{VEV}s]{VEV}{VEV}{vacuum expectation value}
\newcommand{\vev}{\gls{VEV}}
\newcommand{\vevs}{\glspl{VEV}}
\newacronym{QM}{QM}{quantum mechanics}

\newacronym[\glslongpluralkey={quantum field theories},%
  shortplural=\abbrev{QFT}s]{QFT}{QFT}{quantum field theory}

\newacronym{UV}{UV}{ultraviolet}
\newcommand{\uv}{\gls{UV}}
\newacronym{QCD}{QCD}{quantum chromodynamics}
\newcommand{\qcd}{\gls{QCD}}
\newacronym{LHC}{LHC}{Large Hadron Collider}

\newacronym{PMNS}{PMNS}{Pontecorvo–Maki–Nakagawa–Sakata}

\newacronym{CKM}{CKM}{Cabbibo-Kobayashi-Maskawa}

\newacronym{IbP}{IbP}{integration-by-parts}

\newacronym{LO}{LO}{leading order}
\newcommand{\lo}{\gls{LO}}
\newacronym{NLO}{NLO}{next-to-leading order}
\newcommand{\nlo}{\gls{NLO}}
\newacronym{NNLO}{NNLO}{next-to-next-to-leading order}
\newcommand{\nnlo}{\gls{NNLO}}
\newacronym{LL}{LL}{leading logarithmic}
\newacronym{NLL}{NLL}{next-to-leading logarithmic}
\newacronym{NNLL}{NNLL}{next-to-next-to-leading logarithmic}

\newacronym[\glslongpluralkey={parton density functions},%
  shortplural=\abbrev{PDF}s]{PDF}{PDF}{parton density function}
\newcommand{\pdf}{\gls{PDF}}
\newcommand{\pdfs}{\glspl{PDF}}
\newacronym{SM}{SM}{Standard Model}

\newacronym{SMEFT}{SMEFT}{Standard Model Effective Field Theory}

\newacronym{BSM}{BSM}{beyond-the-\gls{SM}}

\newacronym{MSSM}{MSSM}{Minimal Supersymmetric \gls{SM}}

\newacronym{SUSY}{SUSY}{Supersymmetry}

\newacronym{DREG}{DREG}{Dimensional Regularization}

\newacronym{DRED}{DRED}{Dimensional Reduction}

\newacronym{EMT}{EMT}{energy-momentum tensor}

\title{Recent applications of the gradient flow}

\author*[a]{Robert Harlander}
\author[a,b,c]{Andrea Shindler}

\affiliation[a]{RWTH Aachen University, Sommerfeldstr.~16, 52074 Aachen,
  Germany} \affiliation[b]{Department of Physics, University of California,
  Berkeley, California 94720, USA} \affiliation[c]{Nuclear Science Division,
  Lawrence Berkeley National Laboratory, Berkeley, California 94720, USA}
\emailAdd{harlander@physik.rwth-aachen.de}
\emailAdd{shindler@physik.rwth-aachen.de}

\abstract{The gradient flow provides a gauge- and O($4$)-invariant
  ultraviolet regulator for \abbrev{QCD}. This opens new opportunities for
  combining non-perturbative lattice calculations with perturbative results,
  in particular in the context of an \ope. We review the concepts behind this
  and argue that, within the operator basis of a given \ope, the flow-time
  dependence of the conversion between flowed and regular Wilson coefficients
  is exact at each perturbative order. Recent applications to parton
  distribution functions, heavy-meson observables, quark masses, and
  gradient-flow definitions of the strong coupling are discussed. Finally,
  the perturbative formulation of the Ricci flow in quantum gravity and its
  application to the study of non-Gaussian fixed points is reviewed.}

\FullConference{Loops and Legs in Quantum Field Theory (LL2026)\\
12-17, April, 2026\\
Bayreuth, Germany\\}

\begin{document}
\maketitle



\glsresetall 

\section{Operator product expansion}\label{sec:ope}

Let us consider the \ope\ of a physical operator $R(Q)$ in momentum space:
\begin{equation}\label{eq::iota}
  \begin{aligned}
    R(Q) &= \sum_{n=1}^N C^\bare_n(Q)\op^\bare_n +
    \order{1/Q^{d_\text{max}+1}} \equiv
    C^\bare(Q)\op^\bare +
    \order{1/Q^{d_\text{max}+1}}\,.
  \end{aligned}
\end{equation}
As an example for $R(Q)$, one may consider a particular Lorentz structure of a
current-current operator. The scale $Q$ is assumed to be large compared to any
other mass scales, and w.l.o.g., we can also assume $R(Q)$ to be
dimensionless; this implies that the sum involves operators of mass dimension
$d\leq d_\text{max}$, and the bare Wilson coefficients $C^\bare_n$ are
proportional to $1/Q^{d_n}$, modulo powers of $\ln Q$, where $d_n$ is the mass
dimension of the operator $\op^\bare_n$.

In general, neither the Wilson coefficients nor matrix elements of the
operators on the r.h.s.\ are \uv\ finite; however, since $R(Q)$ is a physical
operator by assumption, the \uv\ divergences must cancel. They can be absorbed
into the renormalization and mixing of the operators and the corresponding
Wilson coefficients,
\begin{equation}\label{eq::idol}
  \begin{aligned}
    C^\bare(Q) = C(Q) Z\,,\qquad \op^\bare = Z^{-1}\op\,,
  \end{aligned}
\end{equation}
such that the \textit{renormalized} Wilson coefficients $C(Q)$ and matrix
elements of the \textit{renormalized} operators $\op$ are \uv\ finite. Unless
stated otherwise, we will assume the renormalization matrix to be in the
\msbar\ scheme in this paper. It then has the form\footnote{We restrict the
discussion to the physical sector of the operator basis, unless stated
otherwise.}
\begin{equation}\label{eq:sftx:farl}
  \begin{aligned}
    Z = \mathds{1} + \sum_{k=1}^\infty
    \sum_{l=1}^k c_{kl}\frac{\alpha_s^k}{\ep^l}\,,
  \end{aligned}
\end{equation}
where $\ep=(4-D)/2$, with $D$ the space-time dimension adopted in dimensional
regularization.

Both the Wilson coefficients as well as the matrix elements of the composite
operators depend on the renormalization scale $\mu$ in general,
\begin{equation}\label{eq:ope:csel}
  \begin{aligned}
    \mu^2\dderiv{}{}{\mu^2}C(Q) = -C(Q)\gamma\,,\qquad
    \mu^2\dderiv{}{}{\mu^2}\langle\op\rangle = \gamma\langle\op\rangle\,,
  \end{aligned}
\end{equation}
where
\begin{equation}\label{eq:ope:fern}
  \begin{aligned}
    \gamma = \left(\mu^2\dderiv{}{}{\mu^2} Z\right)Z^{-1}
  \end{aligned}
\end{equation}
is the anomalous dimension matrix of the set of operators $\op$. In matrix
elements of the physical operator 
\begin{equation}\label{eq:ope:rms}
  \begin{aligned}
    R(Q)= C(Q)\op + \order{1/Q^{d_\text{max}+1}}\,,
  \end{aligned}
\end{equation}
the $\mu$-dependence cancels exactly; in perturbation theory, a residual scale
dependence remains which is formally of higher order.

Typically, $C(Q)$ is computed perturbatively, while the hadronic matrix
elements of $\op$ account for the non-perturbative effects. Their combination
requires the lattice operators to be renormalized in a suitable scheme, such
as \abbrev{RI/MOM}~\cite{Martinelli:1994ty},
\abbrev{RI/SMOM}~\cite{Aoki:2007xm,Sturm:2009kb}, or the Schr\"odinger
functional (\abbrev{SF})
scheme~\cite{Luscher:1991wu,Luscher:1992an,Sint:1993un}, and subsequently
matched to the scheme used for the Wilson coefficients.  This requires
dedicated simulations, typically involving off-shell kinematical
configurations in the former cases, or finite-volume simulations in the
latter.

The calculation of matrix elements of composite operators on the lattice faces
a number of conceptual difficulties though.  In particular, since the lattice
introduces the dimensionful cutoff $a^{-1}$, \uv\ divergences can appear as
inverse powers of the lattice spacing $a$, leading to so-called power
divergences which can be difficult, if not impossible, to control
numerically. Additional conceptual problems arise if the operators carry
Lorentz indices. The lattice explicitly breaks the O($4$) symmetry of
continuous Euclidean space-time to the hypercubic group H($4$), so that \uv\
divergences are classified according to irreducible representations of
H($4$). As a consequence, operators belonging to different O($4$) irreducible
representations may mix if they belong to the same H($4$) representation,
including mixing with lower-dimensional operators.  The gradient flow provides
a gauge- and O($4$)-invariant continuum regulator for short-distance
singularities in composite operators which can be implemented both
perturbatively and non-perturbatively on the lattice.


\section{Gradient flow}\label{sec:gf}

In order to motivate the gradient flow, let us begin by naively removing the
high-momentum modes of the fields in the operator in a Wilsonian fashion. For
the gauge field of \qcd, we would write this as (we adopt a Euclidean metric
throughout this presentation, unless stated otherwise)
\begin{equation}\label{eq::egad}
  \begin{aligned}
    A_\mu(x)=  \int\dd^4 p\,
    e^{ipx}\tilde{A}_\mu(p)\quad\to\quad
    A_{\mu}(\Lambda,x)=  \int\dd^4 p\,
    \theta(\Lambda^2-p^2)e^{ipx}\tilde{A}_\mu(p)\,.
  \end{aligned}
\end{equation}
This cutoff preserves translations and O($4$), but its coordinate-space kernel
is nonlocal and has power-law tails. Moreover, a naive momentum-space cutoff
on the gauge field does not admit a straightforward gauge-covariant
generalization to the interacting non-Abelian theory. Instead of the $\theta$
function, we could therefore introduce a softer momentum cutoff as
\begin{equation}\label{eq::fixt}
  \begin{aligned}
    A_\mu(t,x) &= \int\dd^4 p\,e^{-tp^2}\,e^{ipx}\tilde{A}_\mu(p)
    =
    \int\frac{\dd^4 y}{(4\pi t)^2}
    \,\exp\left[-\frac{(x-y)^2}{4t}\right]A_\mu(y)\,,
  \end{aligned}
\end{equation}
corresponding to Gaussian smearing of the unregularized field with smearing
radius $\sqrt{8t}$.  Obviously, this field is the unique solution of
\begin{equation}\label{eq::ilya}
  \begin{aligned}
    \partial_tA_\mu(t,x) &= \Box A_\mu(t,x)\,,\qquad A_\mu(t=0,x)=A_\mu(x)\,,
  \end{aligned}
\end{equation}
which is O($4$) invariant in Euclidean space: if $A_\mu(t,x')$ is a solution,
then so is the transformed field $A'_\mu(t,x)$, where $x$ and $x'$ as well as
$A$ and $A'$ are related by an O($4$) rotation. Requiring $A_\mu(t,x)$ to
transform in the same way under gauge transformations as $A_\mu(x)$,
\cref{eq::ilya} is obviously not gauge covariant. However, due to the $t$
derivative, the \lhs\ transforms homogeneously,
\begin{equation}\label{eq::biro}
  \begin{aligned}
    \partial_t A_\mu(t,x)\to U^\dagger(x)\partial_t
    A_\mu(t,x)U(x)\,, 
  \end{aligned}
\end{equation}
where $U(x)\in \text{SU}(3)$. A gauge-covariant generalization is provided by
the \qcd\ gradient-flow
equation~\cite{Narayanan:2006rf,Luscher:2009eq,Luscher:2010iy}:
  \begin{equation}\label{eq::halm}
  \begin{aligned}
    \partial_tB_\mu(t,x) &= D_\nu(t,x) G_{\nu\mu}(t,x)\,,\qquad
    B_\mu(t=0,x)=A_\mu(x)\,,
  \end{aligned}
\end{equation}
where
\begin{equation}\label{eq::adze}
  \begin{aligned}
    D_\mu = \partial_\mu + g_\bare [B_\mu,\cdot]\,,\qquad G_{\mu\nu} =
    \partial_\mu B_\nu - \partial_\nu B_\mu + g_\bare[B_\mu,B_\nu]\,,
  \end{aligned}
\end{equation}
with $g_\bare$ the bare strong coupling. At tree level, and for
$\partial_\mu B_\mu=0$, \cref{eq::halm} reduces to the heat equation in
\cref{eq::ilya}. Analogously, flowed quark fields are
defined through~\cite{Luscher:2013cpa}
\begin{equation}\label{eq:gf:isbn}
  \begin{aligned}
    \partial_t\chi(t,x) = D_\text{F}^2\chi(t,x)\,,\qquad
    \chi(t=0,x)=\psi(x)\,,
  \end{aligned}
\end{equation}
where $D_{\text{F},\mu}=\partial_\mu + g_\bare B_\mu$. 

Correlation functions of operators which are composed of flowed fields,
referred to as \textit{flowed operators} in the following, are \uv\ finite for
positive flow time once the \qcd\ parameters are renormalized in the regular
way~\cite{Luscher:2010iy,Luscher:2011bx}. In the usual lattice normalization,
no additional renormalization of flowed gauge-field observables is required;
in the normalization defined by \cref{eq::halm,eq::adze}, a factor of $Z_g$
needs to account for the explicit factor of $g_\bare$ though.  Flowed fermion
fields, on the other hand, require the separate renormalization constant
$Z_\chi$~\cite{Luscher:2013cpa,Artz:2019bpr}.


\section{Flowed operator product expansion}\label{sec:fope}


\subsection{General considerations}\label{sec:general}

Assume that one can write $R(Q)$ of \cref{eq::iota} in terms of flowed
operators~\cite{Monahan:2015lha,Harlander:2020duo,Beneke:2025hlg}, i.e.
\begin{equation}\label{eq::brad}
  \begin{aligned}
    R(Q) &= \sum_n\tilde{C}_n(Q,t)\tilde{\op}_n(t) +
    \order{1/Q^{d_\text{max}+1}}
    \equiv \tilde{C}(Q,t)\tilde{\op}(t) + 
    \order{1/Q^{d_\text{max}+1}}\,.
  \end{aligned}
\end{equation}
We can determine the relation between the flowed and the unflowed Wilson
coefficients by constructing \textit{projectors} onto the unflowed bare
operators as~\cite{Gorishnii:1983su,Harlander:2018zpi}
\begin{equation}\label{eq::care}
  \begin{aligned}
    P_n[\op^\bare_m] = \delta_{nm}\,.
  \end{aligned}
\end{equation}
These projectors are constructed from suitable matrix elements, potentially
including derivatives w.r.t.\ the external mass scales (masses and momenta),
and evaluated with all of these mass scales set to zero.  It follows that
\begin{equation}\label{eq::krio}
  \begin{aligned}
    \tilde{C}(Q,t) &= C^\bare(Q)\zeta^{\bare}(t)^{-1}\,,
  \end{aligned}
\end{equation}
where we have introduced the bare \textit{matching matrix}
\begin{equation}\label{eq::acle}
  \begin{aligned}
    \zeta^\bare_{mn}(t) &= P_n[\tilde{\op}_m(t)]\,.
  \end{aligned}
\end{equation}
Since $t$ is the only dimensionful scale on the \rhs, it must be
\begin{equation}\label{eq::auge}
  \begin{aligned}
    \zeta^\bare_{mn}(t) \sim t^{(d_n-d_m)/2} \sim
    (\zeta^\bare(t)^{-1})_{mn}\,,
  \end{aligned}
\end{equation}
modulo logarithmic terms in $t$.  If one evaluates the bare matching matrix
via \cref{eq::acle} in dimensional regularization, it contains $1/\ep$ poles
which, however, can be factorized into the regular renormalization matrix $Z$,
introduced in \cref{eq::idol}, as
\begin{equation}\label{eq:fope:clay}
  \begin{aligned}
    \zeta(t) = \zeta^\bare(t) Z^{-1}\quad
    \Rightarrow\quad \tilde{C}(Q,t)=C(Q)\zeta(t)^{-1}\,.
  \end{aligned}
\end{equation}
Note that, up to this point, we have not made any assumptions on the magnitude
of $t$. In a lattice calculation of the flowed matrix elements, the relevant
short-distance scales are $t$ and the lattice spacing $a$. The continuum limit
is taken at fixed positive $t$, while at finite lattice spacing the smearing
radius should be sufficiently large compared with the cutoff scale,
e.g.\ $\sqrt{8t}\gtrsim a_\text{max}$ when several lattice spacings enter the
continuum extrapolation.
Moreover, at fixed perturbative order, $1/\sqrt{t}$ and $Q$ should not be too
different in order to avoid large logarithms in the flowed Wilson
coefficients, thus $t=1/(\kappa Q^2)$, with $\kappa \sim \order{1}$. The
variation of the final result with a variation of $\kappa$ may then serve as
an estimate of the theoretical uncertainty.

It is important to note that, in the context of the \ope\ defined in
\cref{eq::iota}, \textit{the flowed Wilson coefficients defined in
  \cref{eq:fope:clay} are exact functions of $t$ at each order in perturbation
  theory}. Contributions associated with additional powers in $t$ would in
general involve higher-dimensional operators beyond the truncated \ope\ of
\cref{eq::iota}.


\subsection{Adler function}\label{sec:adler}

As an application, consider the \ope\ of the Adler function (neglecting
quark-mass effects; for a precise definition, see \citere{Beneke:2025hlg}, for
example)
\begin{equation}\label{eq::kibe}
  \begin{aligned}
    D(Q) = D_\text{pert}(Q) + C^\bare_G(Q)\,
    \langle F_{\mu\nu}F_{\mu\nu}\rangle_0 + \order{Q^{-5}}\,,
  \end{aligned}
\end{equation}
where $\langle\cdot\rangle_0$ denotes the \vev; i.e., in \cref{eq::brad}, we
identify
\begin{equation}\label{eq::hast}
  \begin{aligned}
    \langle R(s)\rangle_0 = D(Q)\,,\qquad
    C^\bare_1(Q) = D_\text{pert}(Q)\,,\qquad
    C^\bare_2(Q) = C^\bare_G(Q)\,.
  \end{aligned}
\end{equation}
Adopting the strategy outlined in \cref{sec:general}, the flowed versions of
the composite operators appearing in \cref{eq::kibe} can be expressed
as\footnote{The flowed unit operator is the unit operator.}
\begin{equation}\label{eq:gg:joey}
  \begin{aligned}
    \begin{pmatrix}
      \mathds{1}\\
      Z_g^2 G_{\mu\nu}(t)G_{\mu\nu}(t)
    \end{pmatrix}
    =
    \zeta^\bare(t)
    \begin{pmatrix}
      \mathds{1}\\
      \,F_{\mu\nu}F_{\mu\nu}
    \end{pmatrix}\,,\qquad
    \zeta^\bare(t) &= \begin{pmatrix} 1 & 0 \\ \zeta_0(t) & \zeta^\bare_G(t)
    \end{pmatrix}\,.
  \end{aligned}
\end{equation}
Inverting the matching matrix, we find
\begin{equation}\label{eq::clay}
  \begin{aligned}
    \tilde{C}_1(Q,t) &=
    D_\text{pert}(Q) - \zeta_0(t)\,r_G(Q,t)\,,\qquad
    \tilde{C}_2(Q,t) = r_G(Q,t)\,,
  \end{aligned}
\end{equation}
where the ratio
\begin{equation}\label{eq::broz}
  \begin{aligned}
    r_G(Q,t) = \frac{C^\bare_G(Q)}{\zeta^\bare_G(t)}
  \end{aligned}
\end{equation}
is \uv\ finite and known through
$\order{\alpha_s^2}$~\cite{Harlander:1998diss,Harlander:2016vzb,
  Artz:2019bpr,Harlander:2020duo,Bruser:2024zyg}.  Therefore,
\begin{equation}\label{eq::helm}
  \begin{aligned}
    D(Q) = \tilde{C}_1(Q,t) + \tilde{C}_2(Q,t)\frac{1}{g^2}E(t) +
    \order{Q^{-5}}\,,
  \end{aligned}
\end{equation}
where $g=g_\bare/Z_g$ is the renormalized coupling, and
\begin{equation}\label{eq::eely}
  \begin{aligned}
    E(t) = g_\bare^2\langle G_{\mu\nu}(\flowt)G_{\mu\nu}(\flowt)\rangle_0
  \end{aligned}
\end{equation}
is the flowed gauge action density.  As pointed out in
\citere{Beneke:2025hlg}, the \rhs\ of \cref{eq::helm} is free from any
renormalon ambiguity which typically plagues the \rhs\ of \cref{eq::kibe}. In
order to properly take into account non-perturbative effects, a precise
continuum extrapolation of a lattice calculation for $E(t)$ would be desirable
in publicly available form.

An important observation is that, in accordance with \cref{eq::auge},
$\zeta_0(t)\sim t^{-2}$. At finite perturbative order, this short-distance
contribution cannot be expected to cancel exactly against the corresponding
contribution contained in a non-perturbative determination of $E(t)$. In the
finite-flow-time \ope, $1/\sqrt{t}$ plays the role of a factorization scale and
should remain below the hard scale $Q$, while at fixed perturbative order a very
large separation between these scales should be avoided in order to control
logarithms of $tQ^2$. As demonstrated in \citere{Beneke:2025hlg}, there is
indeed a rather broad region of flow times where the residual $t$ dependence is
weak, providing an important check of the finite-flow-time construction.


\subsection{Flow-time evolution}\label{sec:gammaf}

Very often, the typical scales for the operator matrix elements and the Wilson
coefficients are widely separated, so that one cannot avoid large logarithms
in a fixed-order calculation of the Wilson coefficients. This is true both in
the regular as well as the flowed \ope. And similar to the former, it is
possible to resum these logarithms in the latter in terms of a \rg-like
equation~\cite{Harlander:2020duo,Borgulat:2023xml},
\begin{equation}\label{eq:fope:kval}
  \begin{aligned}
    \flowt\partial_\flowt \tilde{C}(Q,\flowt) =
    -\tilde{C}(Q,\flowt)\tilde\gamma(\flowt)\,,\qquad
    \flowt\partial_\flowt \tilde{O}(\flowt) =
    \tilde\gamma(\flowt)\tilde{O}(\flowt)\,,
  \end{aligned}
\end{equation}
where
\begin{equation}\label{eq:fope:ency}
  \begin{aligned}
    \tilde{\gamma}(\flowt) =
    \left(\flowt\dderiv{}{}{\flowt}\zeta(\flowt)\right)\zeta^{-1}(\flowt)\,,
  \end{aligned}
\end{equation}
and we would like to point out the formal resemblance to
\cref{eq:ope:csel,eq:ope:fern}.  This equation can be solved analogously to
the standard \rg\ equation in the \msbar\ scheme.  Note, however, that as
opposed to the latter, \cref{eq:fope:kval} mixes Wilson coefficients
corresponding to different mass dimensions in general, i.e.,
\begin{equation}\label{eq:fope:echo}
  \begin{aligned}
    \tilde{\gamma}_{mn}(\flowt) =
    \sum_k\left(\flowt\dderiv{}{}{\flowt}\zeta_{mk}(\flowt)\right)
    \zeta^{-1}(\flowt)_{kn} \sim t^{(d_n-d_m)/2}\,,
  \end{aligned}
\end{equation}
with the same notation as in \cref{sec:general}. Note that each term in the
sum scales as $t^{(d_n-d_m)/2}$, independent of $k$. Thus, outside of the
truncated \ope\ framework introduced above, the sum over $k$ requires an
infinite number of terms. However, within the \ope\ context, the sum is
limited by the finite size of the operator basis, induced via the requirement
$d_k\leq d_\text{max}$. Interestingly, increasing $d_\text{max}$ thus changes
the elements of $\tilde{\gamma}$; however, it also increases the size of the
operator basis. The change in evolution due to one element $\tilde\gamma_{mn}$
is thus (partly) compensated by the additional operators in the evolution of
the individual $\tilde{C}_n(Q,\flowt)$.

The main message here is that, within the chosen \ope\ operator basis,
$\tilde{\gamma}(\flowt)$ as defined through \cref{eq:fope:kval} is
well-defined and \textit{an exact}\footnote{It still is a perturbative
expression in $\alpha_s$, of course. For a non-perturbative approach to
$\tilde{\gamma}$, see \citere{Black:2026kfy}.} \textit{function of $\flowt$}.


\subsection{Short-flow-time expansion}\label{sec:sftx}

Instead of converting the \msbar\ Wilson coefficients to the flowed scheme as
in the previous sections, one may also apply the matching matrix to the flowed
operators:
\begin{equation}\label{eq:sftx:deft}
  \begin{aligned}
    \op = \zeta^{-1}(t)\tilde{\op}(t)\,.
  \end{aligned}
\end{equation}
In the context of an \ope\ as in \cref{eq::iota}, this is equivalent
to the procedure described in \cref{sec:general}, of course, if we
consistently expand the expression for $R(Q)$ in $\alpha_s$.

The main purpose of \cref{eq:sftx:deft}, though, is to apply it
\textit{outside} the context of an \ope, i.e., without having an external
\uv\ scale $Q$ in mind, and taking the limit $t\to 0$.  In this sense, the
\rhs\ should be considered as reconstructing the \msbar-renormalized operator;
this is the so-called \sftx.  However, both factors on the \rhs\ of
\cref{eq:sftx:deft} contain divergences in this limit. The fact that they are
calculated using different methods (lattice and perturbation theory) requires,
at finite perturbative order and numerical precision, additional efforts to
control the cancellation of these divergences. In the case of purely
logarithmic divergences, this has been
successfully studied in quite some detail, as the examples to be considered
below will show. First investigations for the case of power divergences were
conducted in \citere{Kim:2021qae}.

Recall that in the case of the flowed \ope\ described in
\cref{sec:general,sec:adler}, the strict limit $t\to 0$ is not required, and
thus these practical problems are avoided.



\section{Applications}\label{sec:applications}


\subsection{Parton distribution functions}\label{sec:pdfs}

An obvious goal for any non-perturbative approach to \qcd\ is to describe the
properties of hadrons from first principles. The calculation of hadron masses
was a milestone achievement in lattice gauge theory~\cite{BMW:2008jgk}, which
has since developed into a precision tool for Standard Model
phenomenology~\cite{DallaBrida:2025ymw,
  FlavourLatticeAveragingGroupFLAG:2024oxs}.  The parton structure of hadrons,
however, still needs to be inferred mostly from experimental data.

The Euclidean formulation of lattice gauge theory is in conflict with the
light-cone definition of \pdfs. Alternatively, one may consider their Mellin
moments,
\begin{equation}\label{eq:appl:gybe}
 \langle x^{n}\rangle_{i/h}(\mu)
 = \int_0^1\dd x\,x^n
 \left[f_{i/h}(x,\mu)+(-1)^{n+1}f_{\bar i/h}(x,\mu)\right] ,
\end{equation}
where $f_{i/h}(x,\mu)$ denotes the distribution of a quark of flavor $i$
carrying a fraction $x$ of the momentum of hadron $h$, and
$f_{\bar i/h}(x,\mu)$ the corresponding antiquark distribution, at
factorization scale $\mu$. These moments can be expressed in terms of
hadronic matrix elements of the twist-two operators
\begin{equation}\label{eq:appl:eath}
 \op^{rs}_{\mu_1\cdots\mu_{n+1}}
 =\bar q^r\gamma_{\{\mu_1}D_{\mu_2}\cdots
 D_{\mu_{n+1}\}}q^s-\text{traces}\,,
\end{equation}
where $r,s$ denote quark flavors, $D_\mu$ is the covariant derivative, and
curly brackets denote symmetrization of the Lorentz indices.

Their direct lattice calculation is complicated at higher moments by the
breaking of O($4$) symmetry to the hypercubic group H($4$), which allows
mixing with lower-dimensional operators and hence introduces power divergences.
This has traditionally restricted calculations with local operators to the
lowest moments. \citere{Shindler:2023xpd} proposed to overcome this limitation
using flowed operators: the continuum limit is taken at fixed positive flow
time, where O($4$) symmetry is restored, thereby eliminating the
power-divergent lower-dimensional mixing induced by the reduced lattice
symmetry. The flowed matrix elements can subsequently be matched to the
physical moments.

For non-singlet flavor configurations, the matching coefficients were obtained
through \nlo\ \qcd\ for arbitrary $n$ in \citere{Shindler:2023xpd}, and
through \nnlo\ for
$n\leq5$~\cite{Harlander:2018zpi,Borgulat:2023xml,Harlander:2025qsa}.
Combined with lattice calculations, this method has enabled the determination
of pion \pdf\ moment ratios up to $\langle x^5\rangle$ and the reconstruction
of the pion \pdf~\cite{Francis:2025rya,Francis:2025pgf}.  A first application
of this approach to the gluon momentum fraction $\langle x\rangle_g$ has also
been presented in Ref.~\cite{Edwards:2026ixr}, using the \nnlo\ matching
coefficients of Ref.~\cite{Harlander:2018zpi}. This exploratory calculation
was performed at a single lattice spacing and neglected quark--gluon mixing.


\subsection{Heavy meson mixings and lifetimes}\label{sec:meson}

The physics of $B$ and $D$ mesons provides sensitive probes of physics beyond
the Standard Model. These observables are a classic example of the interplay
between perturbative Wilson coefficients and non-perturbative matrix elements
of local operators, making them a natural application of the gradient-flow
framework discussed above. The non-perturbative contributions are typically
parametrized in terms of so-called bag parameters, which are combined with
suitable perturbative Wilson coefficients to obtain the physical predictions
for lifetimes or mixing parameters.


%
\begin{figure}
  \begin{center}
    \begin{tabular}{c}
      \raisebox{0em}{%
        \mbox{%
          \includegraphics[%
            width=.5\textwidth]%
                          {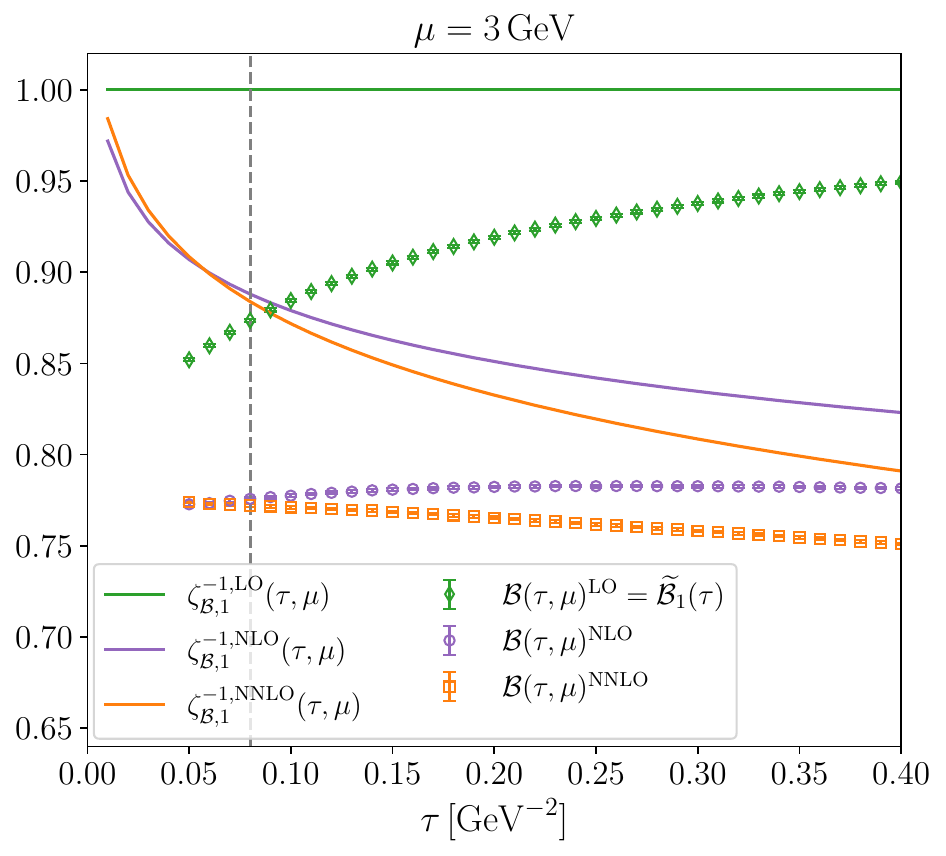}}}
    \end{tabular}
    \parbox{.9\textwidth}{
      \caption[]{\label{fig:mfp}\sloppy The flowed bag parameter
        $\tilde{\mathcal{B}}_1(\flowt)$ (green diamonds), the inverse matching
        coefficient $\zeta^{-1}_{\mathcal{B},1}(\flowt,\mu)$ (solid lines) at
        \lo\ (green), \nlo\ (purple), and \nnlo\ (orange), and their products
        with $\tilde{\mathcal{B}}_1(\flowt)$ (\nlo: purple circles; \nnlo:
        orange squares).  The flow time $\flowt$ is denoted as $\tau$ in these
        plots; the renormalization scale is set to $\mu=3$\,GeV.  }}
  \end{center}
\end{figure}
%


As an example, \cref{fig:mfp} illustrates the extraction of the bag parameter
$\mathcal{B}_1(\mu)$ related to the mixing of a fictitious neutral $D_s$ meson
into its anti-particle using the gradient flow. The flowed bag parameter
$\tilde{\mathcal{B}}_1(t)$ is constructed from matrix elements of flowed four-
and two-quark operators; see \citere{Black:2026rbz,Black:2026dzp} for details.
The relevant point here is that $\tilde{\mathcal{B}}_1(t)$ can be related to
its \msbar\ value by using \cref{eq:sftx:deft},
\begin{equation}\label{eq:appl:krak}
  \begin{aligned}
    \mathcal{B}_1(\mu) = \lim_{\flowt\to 0}\zeta^{-1}_{\mathcal{B},1}(\flowt,\mu)
    \tilde{\mathcal{B}}_1(\flowt)\,,
  \end{aligned}
\end{equation}
where we have made the $\mu$-dependence explicit. For the particular bag
parameter $\mathcal{B}_1$ considered here, the relevant matching relation
reduces to a single coefficient $\zeta_{\mathcal{B},1}(\flowt,\mu)$, known
through \nnlo\ \qcd~\cite{Harlander:2022tgk,Borgulat:2023xml}.
\Cref{fig:mfp} shows the flowed lattice bag parameter, the inverse
perturbative matching coefficient, and their product at \nlo\ and \nnlo.
While the weaker flow-time dependence of the \nlo\ result may appear
preferable at first sight, the leading corrections neglected in the \sftx\ are
parametrized by terms starting linearly in $\flowt$ in the fits. In any case,
both the \nlo\ and the \nnlo\ results allow for a consistent continuum
extrapolation at fixed positive flow time followed by an extrapolation to
$\flowt\to0$; for details, see
\citere{Black:2026rbz,Black:2026dzp}.


\subsection{Quark masses}\label{sec:mass}

An even simpler application which involves only two-quark operators is the
extraction of the quark mass $m(\mu)$ in the \msbar\ scheme from lattice
data. Starting from the \abbrev{PCAC} relation and taking the large
Euclidean-time, ground-state limit, one derives
\begin{equation}\label{eq:appl:gong}
  \begin{aligned}
    \frac{M_\pi}{2}\frac{\langle A_4(x_4)P(0)\rangle_0}{\langle
      P(x_4)P(0)\rangle_0} = m_\bare^\text{\abbrev{PCAC}}\,,
  \end{aligned}
\end{equation}
where $P(x_4)$ and $A_\mu(x_4)$ are the bare non-singlet zero-momentum
pseudoscalar density and axial current at Euclidean time $x_4$, $M_\pi$ is
the pion mass, and $m_\bare^\text{\abbrev{PCAC}}$ is the bare PCAC mass.
Replacing these operators by their flowed counterparts in the corresponding
ratio, where the flowed-quark field renormalization cancels, defines a
\uv-finite gradient-flow version of the quark mass, which is related to the
\msbar\ mass as
\begin{equation}\label{eq:appl:bess}
  \begin{aligned}
    m(\mu) = \lim_{\flowt\to 0}
    \zeta_P(\flowt,\mu)\zeta_A(\flowt,\mu)^{-1}\tilde{m}(\flowt)\,,
  \end{aligned}
\end{equation}
where $\zeta_A$ and $\zeta_P$ are the matching coefficients for the
axial current and pseudoscalar density, renormalized in the \msbar\ scheme. The
method has been applied successfully to the strange- and charm-quark masses in
\citere{Black:2025gft}. For an alternative way to evaluate \msbar\ quark
masses from the lattice using gradient flow observables,
see~\citere{Takaura:2025pao,Harlander:2025jrm}.



\section{Gradient flow coupling}\label{sec:gg}


\subsection{Quantum chromodynamics}\label{sec:qcd}

The flowed gauge action density introduced in \cref{eq::eely} has been a
central object for applications of the gradient flow ever since the seminal
paper on the gradient flow by M.~Lüscher~\cite{Luscher:2010iy}. Aside from its
role in scale setting on the lattice, it provides a renormalization scheme for
the strong coupling which is accessible perturbatively and on the lattice:
\begin{equation}\label{eq:gg:gait}
  \begin{aligned}
    \tilde{\alpha}_s(\mu) &= \frac{4\pi}{3}t^2E(t)\Big|_{t=c/\mu^2}\,.
  \end{aligned}
\end{equation}
The parameter $c$ is in principle arbitrary, but is typically set to
$1/8$, corresponding to $\mu=1/\sqrt{8t}$. The perturbative conversion to the
\msbar\ scheme is known through \nnlo\ and shows good perturbative behavior
down to scales of order
$1$\,GeV~\cite{Harlander:2016vzb,Artz:2019bpr,Harlander:2025jrm}.\footnote{A
finite-volume evaluation of $E(t)$ using numerical stochastic perturbation
theory through \nnlo\ was obtained in \citere{DallaBrida:2017tru}.} Comparing
to a continuum-extrapolated lattice evaluation of $E(t)$ with $n_f=3$ flavors,
\citere{Larsen:2025wvg} derives the \msbar\ value
$\alpha^{(5)}_s(M_Z)=0.1161^{+0.0027}_{-0.0016}$, compatible with the world
average of this quantity, with a somewhat larger uncertainty.

Interestingly, the first non-universal coefficient of the $\beta$ function in
the gradient-flow scheme is considerably larger than in the
\msbar\ scheme~\cite{DallaBrida:2019wur,Harlander:2021esn,Hasenfratz:2023bok}.
This suggests a more limited perturbative regime for extractions based on
this $\beta$ function, and may require lattice data down to very small values of
$\flowt$~\cite{Hasenfratz:2023bok}. It also motivates the calculation of the
next perturbative term which, however, corresponds to a four-loop calculation.

\begin{figure}
  \begin{center}
    \begin{tabular}{c}
      \raisebox{0em}{%
        \mbox{%
          \includegraphics[%
            width=.5\textwidth]%
                          {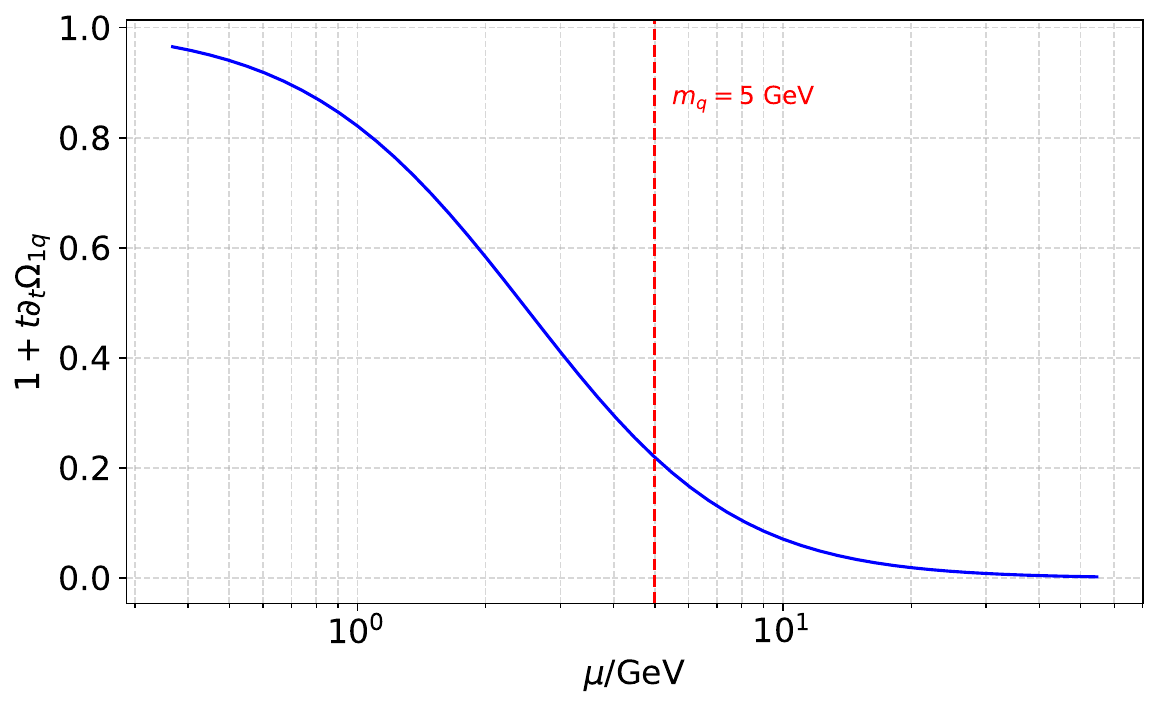}}}
    \end{tabular}
    \parbox{.9\textwidth}{
      \caption[]{\label{fig:beta0}\sloppy
        The function $1+\flowt\partial_\flowt\Omega_{1q}$ which governs the 
        quark contributions to the one-loop $\beta$ function in the
        \gf\ scheme. Here, $\flowt=1/(8\mu^2)$, and $m_q=5$\,GeV.
    }}
  \end{center}
\end{figure}

As opposed to the \msbar\ scheme which is mass independent and requires
explicit matching relations for the transition from \qcd\ with $n_f$ flavors
to $n_f\pm1$ flavors, the gradient-flow scheme implements this decoupling in a
natural way. Using the results of \citere{Harlander:2016vzb} which include the
quark mass effects in $E(t)$ through \nlo, one finds for the one-loop
coefficient of the $\beta$ function\footnote{Thanks to H.~Werthenbach for
motivating this study. The analysis could also be taken to the two-loop level
using the results of \citere{Harlander:2025jrm}.}
\begin{equation}\label{eq:gg:inyl}
  \begin{aligned}
    \tilde{\beta}_0 = 11 - \frac{2}{3}\sum_q\left(1 +
    \flowt\partial_\flowt\Omega_{1q}\right)_{\flowt=c/\mu^2}\,,
  \end{aligned}
\end{equation}
where $\Omega_{1q}$ is a function of the dimensionless combination
$m_q^2\flowt$. The term in brackets of \cref{eq:gg:inyl} is shown in
\cref{fig:beta0} for $m_q=5$\,GeV; from \citere{Harlander:2016vzb}, one
derives that it obeys the following limits:
\begin{equation}\label{eq:gg:echt}
  \begin{aligned}
    1+\flowt\partial_\flowt\Omega_{1q} \to \left\{
    \begin{array}{rl}
      1 &\text{for}\ m_q\to 0\,,\\
      0 &\text{for}\ m_q\to \infty\,.
    \end{array}
    \right.
  \end{aligned}
\end{equation}
Thus, in the case of $n_f$ massless and otherwise only infinitely heavy
quarks, it is $\tilde{\beta}_0=11-2n_f/3$, in agreement with the universal
result for mass independent schemes. As one lowers one of the heavy quark
masses $m_q$, $\tilde{\beta}_0$ continuously decreases until it reaches
$\tilde{\beta}_0=11-2(n_f+1)/3$ at $m_q=0$.  Let us point out that an
alternative perturbative definition of the \qcd\ $\beta$ function which
interpolates smoothly across the thresholds has recently been introduced in
\citere{Kluth:2024lar}.

The fact that the \gf\ scheme is mass dependent has another interesting
consequence. While the \rg\ evolution in the \msbar\ scheme is governed by
logarithmic divergences only, so that power-law sensitivity to dimensionful
parameters is not manifest in the running, the mass-dependent \gf\ scheme also
retains sensitivity to such power-law effects. This will become relevant in
the subsequent discussion.


%
%



\subsection{Quantum electrodynamics}\label{sec:qed}

As mentioned in \cref{sec:qcd}, the perturbative behavior of the first
non-universal coefficient of the \gf\ $\beta$ function motivates further
investigation. One option is to consider the case of \QED, for which the flow
equation is linear in the gauge field and can be solved exactly in momentum
space. Because of this, the flowed \QED\ action density can be expressed as a
Laplace transform of the ordinary photon vacuum-polarization
function~\cite{Georg:2026ozz}. Since the latter is known to four-loop level in
\QED, it is possible to evaluate the \QED\ coupling in the \gf\ scheme to the
same order (which would correspond to a five-loop calculation performed
directly in the gradient-flow formalism). It turns out that the perturbative
behavior of the corresponding $\beta$ function is comparable to that in the
\msbar\ scheme, which suggests that the features observed in \qcd\ are of
non-Abelian origin.

Another interesting application of the \QED\ \gf\ coupling is to study the
\rg\ structure of \QED$_3$, i.e., \QED\ in three space-time dimensions. It is
well known that \QED$_3$ shares several qualitative features with
four-dimensional \qcd, including strong infrared dynamics and, for suitable
$n_f$, chiral symmetry breaking. The gradient flow provides an independent
tool to study such features.  For example, in the large-$n_f$ limit it
reproduces both the ultraviolet and infrared fixed points of
\QED$_3$~\cite{Georg:2026ozz}. Non-perturbative investigations of this system
could provide further insight in this respect.



\section{Ricci flow}\label{sec:ricci}

As already pointed out in \cref{sec:qcd}, one of the virtues of the
\gf\ scheme compared to the \msbar\ scheme is the scale-dependent
renormalization condition which makes the \rg\ structure sensitive to
power-law effects associated with dimensionful parameters. It was pointed out in
\citere{Kluth:2024lar} that this is crucial in the context of gravity, where
non-perturbative studies provide evidence for a non-Gaussian fixed point
underlying the asymptotic-safety scenario for quantum gravity (see, e.g.,
\citere{Falls:2014tra}). At the same time, \citere{Kluth:2024lar} introduced a
non-minimal generalization of the \msbar\ scheme which preserves sensitivity
to such power-law effects and indeed recovers a non-Gaussian fixed point.

From the discussion of the gradient flow in \qcd, it is suggestive to
introduce a \gf-like coupling in gravity and to study its associated
\rg\ behavior. For the sake of simplicity, we only consider pure gravity
without any matter fields. The only dynamical field in this case is the metric
tensor, and it is natural to define a flow equation analogous to
\cref{eq::halm} as
\begin{equation}\label{eq:ricc:gale}
  \begin{aligned}
    \partial_\flowt \tilde{g}_{\mu\nu}(\flowt) = -2\tilde{R}_{\mu\nu}(\flowt)\,,
  \end{aligned}
\end{equation}
where $\tilde{R}_{\mu\nu}(\flowt)$ is the flowed version of the Ricci
tensor. With the conventional negative sign, the linearized flow around flat
space has a diffusive character and exponentially suppresses high-momentum
modes. The factor~2 is the conventional normalization of the Ricci-flow
equation introduced by Hamilton~\cite{Hamilton:1982} and later brought to fame
by Perelman as a tool to prove the Poincar\'e
conjecture~\cite{Perelman:2003uq,Perelman:2006un,Perelman:2006up}.

A perturbative formulation of the Ricci flow in quantum gravity was recently
developed in \citere{Harlander:2026aav}. The metric, both flowed and unflowed,
was expanded around a constant, flat background,
\begin{equation}\label{eq:ricc:dock}
  \begin{aligned}
    g_{\mu\nu}(x) = \delta_{\mu\nu} + h_{\mu\nu}(x)\,,\qquad
    \tilde{g}_{\mu\nu}(t,x) = \delta_{\mu\nu} + \tilde{h}_{\mu\nu}(t,x)\,,
  \end{aligned}
\end{equation}
with $h_{\mu\nu}$, $\tilde{h}_{\mu\nu}$ the unflowed and flowed graviton
fields. The unflowed Feynman rules of quantum gravity were supplemented by the
flowed graviton propagator, graviton flow lines, and the corresponding vertices,
taking into account suitable gauge fixing terms for the flowed graviton.

Since Newton's coupling $G_\text{N}$ is dimensionful, perturbative Einstein
gravity is non-\-re\-nor\-mal\-izable by power counting and requires
higher-dimensional counterterms. At \nlo, these may be represented by terms of
the form $R^2$ and $R_{\mu\nu}R^{\mu\nu}$. In flowed gravity, the
corresponding counterterms receive contributions from the \uv\ divergences of
the unflowed Green's functions, as well as from divergences associated with
the flowed fields at vanishing flow time. Once determined, they allow one to
evaluate \vevs\ of composite operators, for example, in analogy to what is
used in \qcd\ in order to define the coupling or the flowed quark
mass~\cite{Luscher:2010iy,Artz:2019bpr,Takaura:2025pao}. In order to arrive at
diffeomorphism invariant quantities, it was suggested to integrate these
quantities over all space:
\begin{equation}\label{eq:ricc:fuss}
  \begin{aligned}
    \mathcal{I}_n(\flowt) = \frac{1}{V}\int\dd^Dx\sqrt{\tilde{g}(\flowt,x)}
    \tilde{O}_n(\flowt,x)\,,
  \end{aligned}
\end{equation}
where $V$ denotes the classical reference volume. \citere{Harlander:2026aav}
considered the simplest two cases, $\tilde{O}_1=\mathds{1}$ and $\tilde{O}_R =
\tilde{R}$. Both $\mathcal{I}_1$ and $\mathcal{I}_R$ were found to be finite
at two-loop level by the same one-loop counterterms. In the \msbar\ scheme,
they are not \rg\ invariant though, and their perturbative results depend on
the renormalization scale $\mu$. This is problematic if one aims for a
comparison to a future non-perturbative implementation of the \rf, for
example. A similar situation exists in \qcd\ due to the requirement of a
flowed-quark field renormalization. It was resolved by defining the so-called
ringed scheme~\cite{Makino:2014taa}, which fixes this renormalization in terms
of a specific \vev.

Analogously, \citere{Harlander:2026aav} suggested to fix the gravity
counterterms such that $\mathcal{I}_1$ does not receive higher-order
corrections. Since $V\mathcal{I}_1$ is the volume of the $D$-dimensional
quantum system, this scheme was dubbed the \textit{\fvs}. In the \fvs,
$\mathcal{I}_1$ is \rg\ invariant by construction, but also $\mathcal{I}_R$
becomes $\mu$ independent. In analogy to \cref{eq:gg:gait}, a possible
definition of Newton's coupling in the \textit{\rf\ scheme} is then
\begin{equation}\label{eq:ricc:arri}
  \begin{aligned}
    g_\text{\rf}(\mu) =
    -\frac{4t}{3}\mathcal{I}_R^\text{\fvs}\Big|_{t=e^{-\EulerGamma}/\mu^2}\,.
  \end{aligned}
\end{equation}
It turns out that this choice indicates both a perturbatively valid conversion
between $g_\text{\rf}(\mu)$ and $G_\text{N}$ as well as a perturbatively
well-behaved $\beta$ function. At the perturbative order considered, the
latter exhibits a \uv\ fixed point at $g_\text{\rf}=4/5$, with associated
critical exponent $\theta=2$, close to values derived from functional
renormalization-group methods~\cite{Baldazzi:2023pep}.

It remains to be seen whether the observations within \rf\ gravity remain
stable against higher-order perturbative corrections, which, however, turn out
to be much more difficult to obtain than in flowed \qcd. Furthermore, it might
be worth implementing the Ricci flow in non-perturbative approaches to quantum
gravity such as Causal Dynamical Triangulation or Euclidean Dynamical
Triangulation,\footnote{For reviews, see
\citeres{Ambjorn:2022naa,Ambjorn:2026prt}, for example.}
providing an independent test of the perturbative result.


\section{Conclusions}

We have collected a personally biased subset of recent developments in the
gradient flow, reflecting its increasing range of applications. Due to space
limitations, some important recent results had to be left out, in particular
the \nlo\ matching of \textit{low-energy effective field theory
  (\abbrev{LEFT})} operators to their flowed
counterparts~\cite{Crosas:2026ofx}, or the application of the gradient flow to
scalar \qcd~\cite{Borgulat:2025gys}. We reviewed the concept of a flowed
\ope~\cite{Monahan:2015lha,Beneke:2025hlg}, where the flow time introduces an
additional scale which is kept at finite positive values, and the strict
$t\to0$ limit is not required. Within the truncated \ope, the resulting flowed
Wilson coefficients are exact functions of the flow time at each perturbative
order, and their flow-time evolution is governed by the flowed anomalous
dimension.  Applications to parton distributions, heavy-meson observables,
quark masses, and gradient-flow couplings illustrate the versatility of this
framework.  Finally, we discussed the Ricci flow as a natural gravitational
analogue of the gradient flow and its recent perturbative application to the
study of a non-Gaussian fixed point in quantum gravity.


\paragraph{Acknowledgments.} We are indebted to all our students,
postdocs, and colleagues for advancing the applications and understanding of
the gradient flow. Some of the insights presented in these proceedings were
inspired by discussions with participants at the 2026 Gradient Flow Workshop
in Edinburgh. R.H.\ would like to thank the organizers of Loops\&Legs for the
invitation to present the gradient flow, and acknowledges support from the
\textit{German Research Foundation (\abbrev{DFG})} through grants~513989149
and 396021762 (\abbrev{TRR}~257 ``Particle Physics Phenomenology after the
Higgs Discovery''); A.S.\ acknowledges support from \textit{National Science
  Foundation} under grant \abbrev{PHY}-2209185.


\bibliographystyle{utphys}
\bibliography{INSPIRE-CiteAll.bib,lit_local.bib}

@inproceedings{Ambjorn:2026prt,
    author = "Ambj{\o}rn, J. and Loll, R.",
    title = "{Causal Dynamical Triangulations: New Lattice Theory of Quantum Gravity}",
    eprint = "2604.05641",
    archivePrefix = "arXiv",
    primaryClass = "hep-th",
    month = "4",
    year = "2026"
}

@article{Baldazzi:2023pep,
    author = "Baldazzi, Alessio and Falls, Kevin and Kluth, Yannick and Knorr, Benjamin",
    title = "{Robustness of the derivative expansion in asymptotic safety}",
    eprint = "2312.03831",
    archivePrefix = "arXiv",
    primaryClass = "hep-th",
    reportNumber = "NORDITA 2023-075",
    doi = "10.1103/hlrm-d4g2",
    journal = "Phys. Rev. D",
    volume = "113",
    number = "2",
    pages = "026005",
    year = "2026"
}

@inbook{Ambjorn:2022naa,
    author = "Ambj{\o}rn, Jan",
    title = "{Lattice Quantum Gravity: EDT and CDT}",
    eprint = "2209.06555",
    archivePrefix = "arXiv",
    primaryClass = "hep-lat",
    doi = "10.1007/978-981-19-3079-9_84-1",
    year = "2024"
}

@article{DallaBrida:2025ymw,
    author = {Dalla Brida, Mattia and H{\"o}llwieser, Roman and Knechtli, Francesco and Korzec, Tomasz and Ramos, Alberto and Sint, Stefan and Sommer, Rainer},
    collaboration = "ALPHA",
    title = "{The strength of the interaction between quarks and gluons}",
    eprint = "2501.06633",
    archivePrefix = "arXiv",
    primaryClass = "hep-ph",
    doi = "10.1038/s41586-026-10339-4",
    journal = "Nature",
    volume = "652",
    number = "8109",
    pages = "328--334",
    year = "2026"
}

@article{FlavourLatticeAveragingGroupFLAG:2024oxs,
    author = "Aoki, Y. and others",
    collaboration = "Flavour Lattice Averaging Group (FLAG)",
    title = "{FLAG review 2024}",
    eprint = "2411.04268",
    archivePrefix = "arXiv",
    primaryClass = "hep-lat",
    reportNumber = "CERN-TH-2024-192, FERMILAB-PUB-24-0785-T",
    doi = "10.1103/nfzp-p5dn",
    journal = "Phys. Rev. D",
    volume = "113",
    number = "1",
    pages = "014508",
    year = "2026"
}

@article{Luscher:1991wu,
    author = "L{\"u}scher, Martin and Weisz, Peter and Wolff, Ulli",
    title = "{A Numerical method to compute the running coupling in asymptotically free theories}",
    reportNumber = "CERN-TH-6008-91",
    doi = "10.1016/0550-3213(91)90298-C",
    journal = "Nucl. Phys. B",
    volume = "359",
    pages = "221--243",
    year = "1991"
}

@article{Luscher:1992an,
    author = "L{\"u}scher, Martin and Narayanan, Rajamani and Weisz, Peter and Wolff, Ulli",
    title = "{The Schr{\"o}dinger functional: A renormalizable probe for non-Abelian gauge theories}",
    eprint = "hep-lat/9207009",
    archivePrefix = "arXiv",
    reportNumber = "DESY-92-025, CERN-TH-6410-92",
    doi = "10.1016/0550-3213(92)90466-O",
    journal = "Nucl. Phys. B",
    volume = "384",
    pages = "168--228",
    year = "1992"
}

@article{Sint:1993un,
    author = "Sint, Stefan",
    title = "{On the Schr{\"o}dinger functional in QCD}",
    eprint = "hep-lat/9312079",
    archivePrefix = "arXiv",
    reportNumber = "DESY-93-165, DESY-93--165",
    doi = "10.1016/0550-3213(94)90228-3",
    journal = "Nucl. Phys. B",
    volume = "421",
    pages = "135--158",
    year = "1994"
}

@article{Black:2026kfy,
    author = "Black, Matthew and Hasenfratz, Anna and Witzel, Oliver",
    title = "{Gradient Flow Renormalization Schemes for Composite Fermion Operators}",
    eprint = "2607.00493",
    archivePrefix = "arXiv",
    primaryClass = "hep-lat",
    reportNumber = "SI-HEP-2026-13, P3H-26-050",
    month = "7",
    year = "2026"
}

@article{Harlander:2026aav,
    author = "Harlander, Robert V. and Kluth, Yannick and Kohnen, Jonas T. and Werthenbach, Henry",
    title = "{The perturbative Ricci flow in gravity}",
    eprint = "2604.18678",
    archivePrefix = "arXiv",
    primaryClass = "hep-th",
    month = "4",
    year = "2026"
}

@article{Black:2026rbz,
    author = "Black, Matthew and Harlander, Robert V. and Kohnen, Jonas T. and Lange, Fabian and Rago, Antonio and Shindler, Andrea and Witzel, Oliver",
    title = "{Bag Parameters for Heavy Meson Lifetimes}",
    eprint = "2603.28516",
    archivePrefix = "arXiv",
    primaryClass = "hep-ph",
    reportNumber = "SI-HEP-2026-07, P3H-26-023, ZU-TH 13/26, TTK-26-06",
    month = "3",
    year = "2026"
}

@article{Black:2026dzp,
    author = "Black, Matthew and Harlander, Robert V. and Kohnen, Jonas T. and Lange, Fabian and Rago, Antonio and Shindler, Andrea and Witzel, Oliver",
    title = "{Heavy-Meson Bag Parameters using Gradient Flow}",
    eprint = "2603.28517",
    archivePrefix = "arXiv",
    primaryClass = "hep-lat",
    reportNumber = "SI-HEP-2026-08, P3H-26-024, ZU-TH 14/26, TTK-26-05",
    month = "3",
    year = "2026"
}

@article{Edwards:2026ixr,
    author = "Edwards, Robert and Karpie, Joe and Maio, Lorenzo and Monahan, Christopher J. and Orginos, Kostas and Richards, David and Sturzu, Alexandru M. and Zafeiropoulos, Savvas",
    collaboration = "HadStruc",
    title = "{Accessing the gluon momentum fraction of nucleons through the gradient flow}",
    eprint = "2602.14260",
    archivePrefix = "arXiv",
    primaryClass = "hep-lat",
    doi = "10.1103/3cc3-mnbs",
    journal = "Phys. Rev. D",
    volume = "114",
    number = "1",
    pages = "014516",
    year = "2026"
}

@article{Crosas:2026ofx,
    author = "Crosas, {\`O}scar L. and Stoffer, Peter",
    title = "{One-loop matching of the LEFT to the QCD gradient flow}",
    eprint = "2601.18883",
    archivePrefix = "arXiv",
    primaryClass = "hep-ph",
    reportNumber = "ZU-TH 02/26, ZU-TH 02/26 ZU-TH 02/26",
    doi = "10.1007/JHEP07(2026)047",
    journal = "JHEP",
    volume = "07",
    pages = "047",
    year = "2026"
}

@article{Georg:2026ozz,
    author = "Georg, Lars and Harlander, Robert V. and Mason, Robert H.",
    title = "{The gradient-flow coupling of three-and four-dimensional QED}",
    eprint = "2601.13914",
    archivePrefix = "arXiv",
    primaryClass = "hep-ph",
    reportNumber = "TTK-25-44, P3H-26-006",
    doi = "10.1103/n586-vj2q",
    journal = "Phys. Rev. D",
    volume = "113",
    number = "7",
    pages = "076016",
    year = "2026"
}

@article{Harlander:2025jrm,
    author = "Harlander, Robert V. and Mason, Robert H.",
    title = "{Quark-mass effects in gradient-flow observables through next-to-next-to-leading order in QCD}",
    eprint = "2512.19613",
    archivePrefix = "arXiv",
    primaryClass = "hep-lat",
    reportNumber = "TTK-25-49, P3H-25-116",
    doi = "10.1007/JHEP05(2026)245",
    journal = "JHEP",
    volume = "05",
    pages = "245",
    year = "2026"
}

@article{Harlander:2025qsa,
    author = "Harlander, Robert V. and Kohnen, Jonas T. and Shindler, Andrea",
    title = "{Short-flow-time expansion of non-singlet twist-two operators at next-to-next-to-leading order QCD}",
    eprint = "2511.17145",
    archivePrefix = "arXiv",
    primaryClass = "hep-ph",
    reportNumber = "TTK-25-41, P3H-25-094",
    doi = "10.1140/epjc/s10052-026-15659-3",
    journal = "Eur. Phys. J. C",
    volume = "86",
    number = "4",
    pages = "439",
    year = "2026"
}

@article{Francis:2025pgf,
    author = "Francis, Anthony and Fritzsch, Patrick and Karur, Rohith and Kim, Jangho and Pederiva, Giovanni and Pefkou, Dimitra A. and Rago, Antonio and Shindler, Andrea and Walker-Loud, Andr{\'e} and Zafeiropoulos, Savvas",
    title = "{Moments of parton distribution functions of the pion from lattice QCD using gradient flow}",
    eprint = "2510.26738",
    archivePrefix = "arXiv",
    primaryClass = "hep-lat",
    doi = "10.1103/vw4v-nyvw",
    journal = "Phys. Rev. D",
    volume = "113",
    number = "7",
    pages = "074520",
    year = "2026"
}

@article{Beneke:2025hlg,
    author = "Beneke, Martin and Takaura, Hiromasa",
    title = "{Gradient-flowed operator product expansion without IR renormalons}",
    eprint = "2510.12193",
    archivePrefix = "arXiv",
    primaryClass = "hep-ph",
    reportNumber = "TUM-HEP-1575/25, YITP-25-162",
    doi = "10.1007/JHEP03(2026)033",
    journal = "JHEP",
    volume = "03",
    pages = "033",
    year = "2026"
}

@article{Francis:2025rya,
    author = "Francis, Anthony and others",
    title = "{Gradient Flow for Parton Distribution Functions: First Application to the Pion}",
    eprint = "2509.02472",
    archivePrefix = "arXiv",
    primaryClass = "hep-lat",
    doi = "10.1103/z3wr-zk8n",
    journal = "Phys. Rev. Lett.",
    volume = "136",
    number = "17",
    pages = "171903",
    year = "2026"
}

@article{Black:2025gft,
    author = "Black, Matthew and Harlander, Robert V. and Hasenfratz, Anna and Rago, Antonio and Witzel, Oliver",
    title = "{Renormalized quark masses using gradient flow}",
    eprint = "2506.16327",
    archivePrefix = "arXiv",
    primaryClass = "hep-lat",
    reportNumber = "P3H-25-044, SI-HEP-2025-15, TTK-25-16",
    doi = "10.1103/3nvr-pkn1",
    journal = "Phys. Rev. D",
    volume = "113",
    number = "9",
    pages = "094504",
    year = "2026"
}

@article{Takaura:2025pao,
    author = "Takaura, Hiromasa and Harlander, Robert V. and Lange, Fabian",
    title = "{A new approach to quark mass determination using the gradient flow}",
    eprint = "2506.09537",
    archivePrefix = "arXiv",
    primaryClass = "hep-lat",
    reportNumber = "YITP-25-89, P3H-25-037, TTK-25-15, ZU-TH 42/25",
    doi = "10.1007/JHEP09(2025)025",
    journal = "JHEP",
    volume = "09",
    pages = "025",
    year = "2025"
}

@article{Larsen:2025wvg,
    author = "Larsen, Rasmus and Mukherjee, Swagato and Petreczky, Peter and Shu, Hai-Tao and Weber, Johannes Heinrich",
    title = "{Scale Setting and Strong Coupling Determination in the Gradient Flow Scheme for 2+1 Flavor Lattice QCD}",
    eprint = "2502.08061",
    archivePrefix = "arXiv",
    primaryClass = "hep-lat",
    month = "2",
    year = "2025"
}

@article{Borgulat:2025gys,
    author = "Borgulat, Janosh and Felten, Nils and Harlander, Robert and Kohnen, Jonas T.",
    title = "{Two-loop gradient-flow renormalization of scalar QCD}",
    eprint = "2501.07150",
    archivePrefix = "arXiv",
    primaryClass = "hep-ph",
    reportNumber = "TTK-24-057",
    doi = "10.21468/SciPostPhysCore.8.1.032",
    journal = "SciPost Phys. Core",
    volume = "8",
    pages = "032",
    year = "2025"
}

@article{Kluth:2024lar,
    author = "Kluth, Yannick",
    title = "{Fixed points of quantum gravity from dimensional regularization}",
    eprint = "2409.09252",
    archivePrefix = "arXiv",
    primaryClass = "hep-th",
    doi = "10.1103/PhysRevD.111.106010",
    journal = "Phys. Rev. D",
    volume = "111",
    number = "10",
    pages = "106010",
    year = "2025"
}

@article{Bruser:2024zyg,
    author = {Br{\"u}ser, Robin and Hoang, Andr{\'e} H. and Stahlhofen, Maximilian},
    title = "{Three-loop OPE Wilson coefficients of dimension-four operators for (axial-)vector and (pseudo-)scalar current correlators}",
    eprint = "2408.03989",
    archivePrefix = "arXiv",
    primaryClass = "hep-ph",
    reportNumber = "FR-PHENO-2024-007, UWThPh-2024-15",
    doi = "10.1007/JHEP12(2024)103",
    journal = "JHEP",
    volume = "12",
    pages = "103",
    year = "2024"
}

@article{Shindler:2023xpd,
    author = "Shindler, Andrea",
    title = "{Moments of parton distribution functions of any order from lattice QCD}",
    eprint = "2311.18704",
    archivePrefix = "arXiv",
    primaryClass = "hep-lat",
    reportNumber = "TTK-23-31",
    doi = "10.1103/PhysRevD.110.L051503",
    journal = "Phys. Rev. D",
    volume = "110",
    number = "5",
    pages = "L051503",
    year = "2024"
}

@article{Borgulat:2023xml,
    author = "Borgulat, Janosch and Harlander, Robert V. and Kohnen, Jonas T. and Lange, Fabian",
    title = "{Short-flow-time expansion of quark bilinears through next-to-next-to-leading order QCD}",
    eprint = "2311.16799",
    archivePrefix = "arXiv",
    primaryClass = "hep-lat",
    reportNumber = "TTK-23-28, P3H-23-075, ZU-TH 72/23, PSI-PR-23-40",
    doi = "10.1007/JHEP05(2024)179",
    journal = "JHEP",
    volume = "05",
    pages = "179",
    year = "2024"
}

@article{Hasenfratz:2023bok,
    author = "Hasenfratz, Anna and Peterson, Curtis Taylor and van Sickle, Jake and Witzel, Oliver",
    title = "{{\ensuremath{\Lambda}} parameter of the SU(3) Yang-Mills theory from the continuous {\ensuremath{\beta}} function}",
    eprint = "2303.00704",
    archivePrefix = "arXiv",
    primaryClass = "hep-lat",
    reportNumber = "FERMILAB-PUB-23-096-V, SI-HEP-2023-04",
    doi = "10.1103/PhysRevD.108.014502",
    journal = "Phys. Rev. D",
    volume = "108",
    number = "1",
    pages = "014502",
    year = "2023"
}

@article{Harlander:2021esn,
    author = "Harlander, Robert",
    title = "{The gradient flow at higher orders in perturbation theory}",
    eprint = "2111.14376",
    archivePrefix = "arXiv",
    primaryClass = "hep-lat",
    reportNumber = "TTK-21-49",
    doi = "10.22323/1.396.0489",
    journal = "PoS",
    volume = "LATTICE2021",
    pages = "489",
    year = "2022"
}

@article{Kim:2021qae,
    author = "Kim, Jangho and Luu, Thomas and Rizik, Matthew D. and Shindler, Andrea",
    collaboration = "SymLat",
    title = "{Nonperturbative renormalization of the quark chromoelectric dipole moment with the gradient flow: Power divergences}",
    eprint = "2106.07633",
    archivePrefix = "arXiv",
    primaryClass = "hep-lat",
    doi = "10.1103/PhysRevD.104.074516",
    journal = "Phys. Rev. D",
    volume = "104",
    number = "7",
    pages = "074516",
    year = "2021"
}

@article{Harlander:2020duo,
    author = "Harlander, Robert V. and Lange, Fabian and Neumann, Tobias",
    title = "{Hadronic vacuum polarization using gradient flow}",
    eprint = "2007.01057",
    archivePrefix = "arXiv",
    primaryClass = "hep-lat",
    reportNumber = "FERMILAB-PUB-20-249-T, IIT-CAPP-20-01, TTK-20-20",
    doi = "10.1007/JHEP08(2020)109",
    journal = "JHEP",
    volume = "08",
    pages = "109",
    year = "2020"
}

@article{DallaBrida:2019wur,
    author = "Dalla Brida, Mattia and Ramos, Alberto",
    title = "{The gradient flow coupling at high-energy and the scale of SU(3) Yang{\textendash}Mills theory}",
    eprint = "1905.05147",
    archivePrefix = "arXiv",
    primaryClass = "hep-lat",
    doi = "10.1140/epjc/s10052-019-7228-z",
    journal = "Eur. Phys. J. C",
    volume = "79",
    number = "8",
    pages = "720",
    year = "2019"
}

@article{Artz:2019bpr,
    author = "Artz, Johannes and Harlander, Robert V. and Lange, Fabian and Neumann, Tobias and Prausa, Mario",
    title = "{Results and techniques for higher order calculations within the gradient-flow formalism}",
    eprint = "1905.00882",
    archivePrefix = "arXiv",
    primaryClass = "hep-lat",
    reportNumber = "FERMILAB-PUB-19-151-T, FR-PHENO-2019-003, IIT-CAPP-19-02, TTK-19-15",
    doi = "10.1007/JHEP06(2019)121",
    journal = "JHEP",
    volume = "06",
    pages = "121",
    year = "2019",
    note = "[Erratum: JHEP 10, 032 (2019)]"
}

@article{Harlander:2018zpi,
    author = "Harlander, Robert V. and Kluth, Yannick and Lange, Fabian",
    title = "{The two-loop energy{\textendash}momentum tensor within the gradient-flow formalism}",
    eprint = "1808.09837",
    archivePrefix = "arXiv",
    primaryClass = "hep-lat",
    reportNumber = "TTK-18-32",
    doi = "10.1140/epjc/s10052-018-6415-7",
    journal = "Eur. Phys. J. C",
    volume = "78",
    number = "11",
    pages = "944",
    year = "2018",
    note = "[Erratum: Eur.Phys.J.C 79, 858 (2019)]"
}

@article{DallaBrida:2017tru,
    author = {Dalla Brida, Mattia and L{\"u}scher, Martin},
    title = "{SMD-based numerical stochastic perturbation theory}",
    eprint = "1703.04396",
    archivePrefix = "arXiv",
    primaryClass = "hep-lat",
    reportNumber = "CERN-TH-2017-057",
    doi = "10.1140/epjc/s10052-017-4839-0",
    journal = "Eur. Phys. J. C",
    volume = "77",
    number = "5",
    pages = "308",
    year = "2017"
}

@article{Harlander:2016vzb,
    author = "Harlander, Robert V. and Neumann, Tobias",
    title = "{The perturbative QCD gradient flow to three loops}",
    eprint = "1606.03756",
    archivePrefix = "arXiv",
    primaryClass = "hep-ph",
    reportNumber = "TTK-16-19",
    doi = "10.1007/JHEP06(2016)161",
    journal = "JHEP",
    volume = "06",
    pages = "161",
    year = "2016"
}

@article{Monahan:2015lha,
    author = "Monahan, Christopher and Orginos, Kostas",
    title = "{Locally smeared operator product expansions in scalar field theory}",
    eprint = "1501.05348",
    archivePrefix = "arXiv",
    primaryClass = "hep-lat",
    reportNumber = "JLAB-THY-15-2008",
    doi = "10.1103/PhysRevD.91.074513",
    journal = "Phys. Rev. D",
    volume = "91",
    number = "7",
    pages = "074513",
    year = "2015"
}

@article{Falls:2014tra,
    author = "Falls, Kevin and Litim, Daniel F. and Nikolakopoulos, Konstantinos and Rahmede, Christoph",
    title = "{Further evidence for asymptotic safety of quantum gravity}",
    eprint = "1410.4815",
    archivePrefix = "arXiv",
    primaryClass = "hep-th",
    reportNumber = "DO-TH-14-26, KA-TP-2014-30",
    doi = "10.1103/PhysRevD.93.104022",
    journal = "Phys. Rev. D",
    volume = "93",
    number = "10",
    pages = "104022",
    year = "2016"
}

@article{Makino:2014taa,
    author = "Makino, Hiroki and Suzuki, Hiroshi",
    title = "{Lattice energy{\textendash}momentum tensor from the Yang{\textendash}Mills gradient flow{\textemdash}inclusion of fermion fields}",
    eprint = "1403.4772",
    archivePrefix = "arXiv",
    primaryClass = "hep-lat",
    reportNumber = "KYUSHU-HET-140",
    doi = "10.1093/ptep/ptu070",
    journal = "PTEP",
    volume = "2014",
    pages = "063B02",
    year = "2014",
    note = "[Erratum: PTEP 2015, 079202 (2015)]"
}

@article{Luscher:2013cpa,
    author = "L{\"u}scher, Martin",
    title = "{Chiral symmetry and the Yang--Mills gradient flow}",
    eprint = "1302.5246",
    archivePrefix = "arXiv",
    primaryClass = "hep-lat",
    reportNumber = "CERN-PH-TH-2013-023",
    doi = "10.1007/JHEP04(2013)123",
    journal = "JHEP",
    volume = "04",
    pages = "123",
    year = "2013"
}

@article{Luscher:2011bx,
    author = "L{\"u}scher, Martin and Weisz, Peter",
    title = "{Perturbative analysis of the gradient flow in non-abelian gauge theories}",
    eprint = "1101.0963",
    archivePrefix = "arXiv",
    primaryClass = "hep-th",
    reportNumber = "CERN-PH-TH-2010-290, MPP-2010-165",
    doi = "10.1007/JHEP02(2011)051",
    journal = "JHEP",
    volume = "02",
    pages = "051",
    year = "2011"
}

@article{Luscher:2010iy,
    author = {L{\"u}scher, Martin},
    title = "{Properties and uses of the Wilson flow in lattice QCD}",
    eprint = "1006.4518",
    archivePrefix = "arXiv",
    primaryClass = "hep-lat",
    reportNumber = "CERN-PH-TH-2010-143",
    doi = "10.1007/JHEP08(2010)071",
    journal = "JHEP",
    volume = "08",
    pages = "071",
    year = "2010",
    note = "[Erratum: JHEP 03, 092 (2014)]"
}

@article{Sturm:2009kb,
    author = "Sturm, C. and Aoki, Y. and Christ, N. H. and Izubuchi, T. and Sachrajda, C. T. C. and Soni, A.",
    title = "{Renormalization of quark bilinear operators in a momentum-subtraction scheme with a nonexceptional subtraction point}",
    eprint = "0901.2599",
    archivePrefix = "arXiv",
    primaryClass = "hep-ph",
    reportNumber = "CU-TP-1186, KANAZAWA-09-01, RBRC-771, SHEP-09-02",
    doi = "10.1103/PhysRevD.80.014501",
    journal = "Phys. Rev. D",
    volume = "80",
    pages = "014501",
    year = "2009"
}

@article{Luscher:2009eq,
    author = "L{\"u}scher, Martin",
    title = "{Trivializing maps, the Wilson flow and the HMC algorithm}",
    eprint = "0907.5491",
    archivePrefix = "arXiv",
    primaryClass = "hep-lat",
    reportNumber = "CERN-PH-TH-2009-118",
    doi = "10.1007/s00220-009-0953-7",
    journal = "Commun. Math. Phys.",
    volume = "293",
    pages = "899--919",
    year = "2010"
}

@article{BMW:2008jgk,
    author = "D{\"u}rr, S. and others",
    collaboration = "BMW",
    title = "{Ab-Initio Determination of Light Hadron Masses}",
    eprint = "0906.3599",
    archivePrefix = "arXiv",
    primaryClass = "hep-lat",
    doi = "10.1126/science.1163233",
    journal = "Science",
    volume = "322",
    pages = "1224--1227",
    year = "2008"
}

@article{Aoki:2007xm,
    author = "Aoki, Y. and others",
    title = "{Nonperturbative renormalization of quark bilinear operators and $B_K$ using domain wall fermions}",
    eprint = "0712.1061",
    archivePrefix = "arXiv",
    primaryClass = "hep-lat",
    reportNumber = "BNL-HET-07-11, CU-TP-1180, EDINBURGH-2007-12, KANAZAWA-07-10, RBRC-681, SHEP-07-20",
    doi = "10.1103/PhysRevD.78.054510",
    journal = "Phys. Rev. D",
    volume = "78",
    pages = "054510",
    year = "2008"
}

@article{Perelman:2006up,
    author = "Perelman, Grisha",
    title = "{Ricci flow with surgery on three-manifolds}",
    eprint = "math/0303109",
    archivePrefix = "arXiv",
    month = "8",
    year = "2006"
}

@article{Perelman:2006un,
    author = "Perelman, Grisha",
    title = "{The Entropy formula for the Ricci flow and its geometric applications}",
    eprint = "math/0211159",
    archivePrefix = "arXiv",
    month = "7",
    year = "2006"
}

@article{Narayanan:2006rf,
    author = "Narayanan, R. and Neuberger, H.",
    title = "{Infinite N phase transitions in continuum Wilson loop operators}",
    eprint = "hep-th/0601210",
    archivePrefix = "arXiv",
    doi = "10.1088/1126-6708/2006/03/064",
    journal = "JHEP",
    volume = "03",
    pages = "064",
    year = "2006"
}

@article{Perelman:2003uq,
    author = "Perelman, Grisha",
    title = "{Finite extinction time for the solutions to the Ricci flow on certain three-manifolds}",
    eprint = "math/0307245",
    archivePrefix = "arXiv",
    month = "7",
    year = "2003"
}

@article{Martinelli:1994ty,
    author = "Martinelli, G. and Pittori, C. and Sachrajda, Christopher T. and Testa, M. and Vladikas, A.",
    title = "{A general method for non-perturbative renormalization of lattice operators}",
    eprint = "hep-lat/9411010",
    archivePrefix = "arXiv",
    reportNumber = "CERN-TH-7342-94, LPTHE-ORSAY-94-52, ROME-1022-1994, SHEP-94-95-03",
    doi = "10.1016/0550-3213(95)00126-D",
    journal = "Nucl. Phys. B",
    volume = "445",
    pages = "81--108",
    year = "1995"
}

@article{Gorishnii:1983su,
    author = "Gorishnii, S. G. and Larin, S. A. and Tkachov, F. V.",
    title = "{The algorithm for OPE coefficient functions in the MS scheme}",
    doi = "10.1016/0370-2693(83)91439-9",
    journal = "Phys. Lett. B",
    volume = "124",
    pages = "217--220",
    year = "1983"
}

@article{Hamilton:1982,
  title = "{Three-manifolds with positive Ricci curvature}",
  author = "Richard S. Hamilton",
  journal = "{Journal of Differential Geometry}",
  year="{1982}",
  volume="{17}",
  pages="{255-306}",
}

@PhdThesis{Harlander:1998diss,
  author    = "Harlander, Robert",
  title     = "{Quarkmasseneffekte in der Quantenchromodynamik
  und asymptotische Entwicklung von Feynman-Integralen}",
  year      = {1998},
  note      = {Shaker Verlag},
  language  = {german},
  publisher = {Shaker Verlag}
}

@article{Harlander:2022tgk,
    author = "Harlander, Robert V. and Lange, Fabian",
    title = "{Effective electroweak Hamiltonian in the gradient-flow formalism}",
    eprint = "2201.08618",
    archivePrefix = "arXiv",
    primaryClass = "hep-lat",
    reportNumber = "TTK-21-58, TTP22-002, P3H-21-104",
    doi = "10.1103/PhysRevD.105.L071504",
    journal = "Phys. Rev. D",
    volume = "105",
    number = "7",
    pages = "L071504",
    year = "2022"
}

\end{document}